\documentclass[a4paper,twocolumn,10pt,accepted=2017-09-16]{quantumarticle}
\pdfoutput=1
\usepackage{amsmath,amssymb}
\usepackage{graphicx}
\usepackage{dcolumn}
\usepackage{bm}
\usepackage{mathrsfs,dsfont}
\usepackage[numbers,sort&compress]{natbib}
\usepackage{hyperref}

\newcommand{\bra}[1]{\langle #1 |} 
\newcommand{\ket}[1]{| #1 \rangle } 
\newcommand{\upd}{\mathrm{d}}

\newcommand{\ie}[0]{\textit{i.e.} }
\newcommand{\eg}[0]{\textit{e.g.} }

\begin{document}


\title{Time-local unraveling of non-Markovian stochastic Schr\"odinger equations}
\date{\today}
\author{Antoine Tilloy}
 \email{antoine.tilloy@mpq.mpg.de}
\affiliation{Max-Planck-Institut f\"ur Quantenoptik, Hans-Kopfermann-Stra{\ss}e 1, 85748 Garching, Germany}
\orcid{0000-0002-6867-9617}

\date{\today}
\begin{abstract}
Non-Markovian stochastic Schr\"odinger equations (NMSSE) are important tools in quantum mechanics, from the theory of open systems to foundations. Yet, in general, they are  but formal objects: their solution can be computed numerically only in some specific cases or perturbatively. This article is focused on the NMSSE themselves rather than on
the open-system evolution they unravel and aims at making them less abstract. Namely, we propose to write the stochastic realizations of linear NMSSE as averages over the solutions of an auxiliary equation with an additional random field. Our method yields a non-perturbative numerical simulation algorithm for generic linear NMSSE that can be made arbitrarily accurate for reasonably short times. For isotropic complex noises, the method extends from linear to non-linear NMSSE and allows to sample the solutions of norm-preserving NMSSE directly. 
\end{abstract}

\maketitle

\section{Introduction}

Stochastic pure state representations have a wide range of applications in quantum mechanics: they serve as computational tools to unravel open-system evolutions \cite{dalibard1992, gisin1992,strunz1999}, as modelling tools to describe continuous measurement situations \cite{jacobs2006,wiseman2008} and as foundational tools to solve the measurement problem in models where the superposition principle breaks down \cite{bassi2003,bassi2013}. In the Markovian limit, the solutions of stochastic Schr\"odinger equations can be efficiently computed numerically, justifying their wide use in the three aforementioned fields. In the non-Markovian case however, there exists no general purpose method to compute a realization of the random pure state process: plotting a single trajectory is in general impossible. In this article, we propose to write the solutions of general stochastic Schr\"odinger equations as averages over the solutions of time-local stochastic Schr\"odinger equations with an auxiliary noise. We shall use the same trick that usually allows to replace openness by stochasticity, this time to get rid of non-Markovianity.

Although stochastic Schr\"odinger equations have a long history dating back to the eighties \cite{gisin1984,diosi1986}, the field took off when their potential in numerics was understood by Dalibard, Castin and M\o lmer \cite{dalibard1992,molmer1993} (see also Dum, Zoller and Ritsch \cite{dum1992}) in the jump case and by Gisin and Percival \cite{gisin1992} in the diffusive case. The connection with actual quantum trajectories coming from realistic measurement setups was understood roughly at the same time by Milburn and Wiseman \cite{wiseman1993} (see \eg \cite{jacobs2006} for an introduction). Interestingly, the introduction of continuous stochastic pure state equations in foundations is slightly anterior with the Continuous Spontaneous Localization (CSL) model of  Pearle, Ghirardi and Rimini \cite{pearle1989,ghirardi1990} and the gravity related collapse model of Di\'osi \cite{diosi1989}, both aimed at solving the measurement problem (see \cite{bassi2003,bassi2013} for a review). 
The generalization of the formalism to the non-Markovian realm was carried out by Di\'osi and Strunz \cite{strunz1996,diosi1997}. In that case, the measurement interpretation exists but is admittedly more subtle: a given trajectory only has a local point by point measurement interpretation \cite{gambetta2002} but no real time meaning \cite{diosi2008,diosi2008erratum,wiseman2008}, at least within orthodox interpretations of the formalism \cite{gambetta2003}. Non-Markovian stochastic approaches have also infected foundations with various extensions of the initial CSL proposal \cite{adler2007,ferialdi2012}. In all these applications, the practical problem is always the same: the equations are very formal, involving functional derivatives, and plotting their solution is typically as hard as solving a general non-Markovian open-system evolution. There are, of course, quite a few cases where the equations can be solved \cite{strunz2004,jing2010,zhao2011,jing2012,jing2013} as well as perturbative methods \cite{devega2005} to deal with a broader class of problems, but general non-perturbative approaches have so far been lacking.

In the following, we will introduce stochastic Schr\"odinger equations as unravelings of non-Markovian open-system dynamics making an extensive use of the general framework introduced recently by Di\'osi and Ferialdi \cite{diosi2014}. Yet, our objective will not primarily be to find a numerically efficient way to compute the open-system evolution: we will be interested in the stochastic pure state trajectories themselves, having in mind their broader range of applications.

\section{General framework }

\subsection{Gaussian master equations}

We consider an open quantum system evolving according to a general Gaussian Master Equation (GME), one of the broadest non-Markovian generalizations of the Lindblad equation that is analytically tractable. We suppose the system has a proper Hamiltonian $H_0$ and will use the interaction picture for all operators $\hat{\mathcal{O}} (t) = e^{i H_0 t} \hat{\mathcal{O}} e^{-i H_0 t}$. The GME for the density matrix in interaction picture can then be written as a time ordered exponential \cite{diosi2014}:
\begin{align}\label{eq:master}
\rho(t) =& \Phi_t \cdot \rho(0)\nonumber \\
=&\mathcal{T} \exp\bigg\{ \int_0^t\!\!\int_0^t \!\! \upd \tau\, \upd s \, D_{ij}(\tau,s) \Big[A^j_L(s) A^i_R(\tau) \\
&- \theta_{\tau s}A^i_L(\tau) A^j_L(s)- \theta_{s\tau}A^j_R(s) A^i_R(\tau) \Big] \bigg\}\,\rho(0) \nonumber
\end{align}
where $\mathcal{T}$ is the time ordering operator, $\{\hat{A}^k\}_{1\leq k\leq n}$ are arbitrary system Hermitian operators, $D_{ij}(\tau,s)$ is a complex positive semi-definite kernel, $\theta_{\tau s}=\theta(\tau-s)$ is the Heaviside function, and we have used summation on repeated indices as well as the left-right notation for superoperators:
\[
\begin{array}{ll}
A_L \cdot \rho = \hat{A}\rho \; ; &A_R \cdot \rho = \rho \hat{A}
\end{array}.
\]
Equation \eqref{eq:master} is a direct non-Markovian generalization of the Lindblad master equation and reduces to it in the limit $D_{ij}(\tau,s)\rightarrow d_{ij}\, \delta(\tau-s)$. It is also an equivalent operator rewriting of the well known quadratic Feynman-Vernon influence functional for systems linearly coupled to harmonic oscillators \cite{feynman1963}, a representation used \eg by Hu, Paz and Zhang \cite{hu1992,hu1993} for the Quantum Brownian Motion and which can be derived from a wide class of microscopic models \cite{weiss1999}. 

At least if $D$ is time translation invariant, equation \eqref{eq:master} can be obtained from the linear coupling of the system of interest with a general bosonic bath (without the Born-Markov approximation). To avoid being excessively abstract, let us recall the explicit derivation of \cite{diosi2014}. We consider a bath constituted of a continuum of frequency of $n$ types of harmonic oscillators. We write the corresponding creation and annihilation operators $a_k(\omega)$, $a_k^\dagger(\omega)$ which verify $[a_i(\omega),a_j^\dagger(\omega')]=\delta_{ij}\delta(\omega-\omega')$. We then couple our system of interest with the bath through the interaction Hamiltonian:
\begin{equation}
H_{\text{int}}=\hat{A}^k \otimes \int\upd \omega\, \kappa_k^\ell a_\ell(\omega)+\kappa_k^{\ell *} a^\dagger_\ell(\omega),
\end{equation}
where $\kappa_k^\ell(\omega)$ is an arbitrary matrix of complex coefficients. If the initial state of the total system is a product with the bath ground state, the reduced system density matrix can be expressed exactly using Wick's theorem \cite{diosi2014} and is given by formula \eqref{eq:master}. In that case, $D$ is given by:
\begin{equation}
D_{ij}(\tau,s)=\int\upd \omega\, \kappa_i^k(\omega) \kappa_j^{\ell *}(\omega) e^{-i\omega(\tau-s)}.
\end{equation}
Here, we are not interested in the open system dynamics \textit{per se} and thus step back from its implementation details to take the very general equation \eqref{eq:master} as our starting point.

\subsection{Stochastic unraveling}

A stochastic unraveling of equation \eqref{eq:master} is a set of pure state trajectories $\ket{\psi_\xi(t)}$ indexed by a set of random fields $\xi=\{\xi_k\}_{1\leq k\leq n}$ such that:
\begin{equation}\label{eq:unraveling}
\rho(t)  = \mathds{E}_\xi \big[\,  \ket{\psi_\xi(t)} \bra{\psi_\xi(t)} \, \big],
\end{equation}
where $\mathds{E}_\xi [\cdot]=\int \cdot\, \upd \mu_\circ(\xi)$ denotes averaging over the set of complex stochastic fields $\xi$.
The interest of this rewriting is that it decouples the left and right parts of the time ordered exponential $ \Phi_t$ of equation \eqref{eq:master}, a trick which is sometimes called a Hubbard-Stratonovich transformation. Let us now find an explicit candidate for $\ket{\psi_\xi(t)}$. Following \eg \cite{diosi2014}, we first introduce a set of Gaussian complex fields of zero average and thus fully characterized by its following two point correlation functions:
\begin{equation}
\begin{split}
    \mathds{E}_\xi\left[\xi_i(\tau)\xi^*_j(s)\right] &= C_{ij}(\tau,s)\\
    \mathds{E}_\xi\big[\xi_i(\tau)\xi_j \,(s)\big] &= S_{ij}(\tau,s)
\end{split}
\end{equation}
where $S$ is the \emph{relation} function, a symmetric complex kernel which can be chosen freely provided the full correlation kernel is positive semi-definite.

\begin{widetext}
\noindent For two sets of fields $a_t^k$ and $b_t^k$, the following generalized characteristic function is obtained by Gaussian integration:
\begin{equation}\label{eq:gaussian}
\mathds{E}_\xi \left[\exp \left\{-i\!\!\int_0^t \!\!\upd s (a^k_s \xi_k(s) -b^k_s \xi_k^*(s))\right\} \right]= \exp\left\{\int_0^t \!\!\int_0^t \!\! \upd \tau \upd s\, C_{ij}(\tau,s) a^i_\tau b^j_s -\frac{S_{ij}(\tau,s) a^i_\tau a^j_s+ S^*_{ij}(\tau,s) b^i_\tau b^j_s }{2}\right\},
\end{equation}
an equality that can be further generalized for $a$'s and $b$'s promoted to operators provided both sides are time ordered: 
\begin{equation}\label{eq:gaussianop}
 \mathds{E}_\xi \left[\mathcal{T}\exp \left\{-i\!\!\int_0^t \!\!\upd s (\hat{a}^k_s \xi_k(s) -\hat{b}^k_s \xi_k^*(s))\right\} \right]= \mathcal{T} \exp\left\{\int_0^t \!\!\int_0^t \!\! \upd \tau \upd s\, C_{ij}(\tau,s) \hat{a}^i_\tau \hat{b}^j_s -\frac{S_{ij}(\tau,s) \hat{a}^i_\tau \hat{a}^j_s+ S^*_{ij}(\tau,s) \hat{b}^i_\tau \hat{b}^j_s }{2}\right\}.
\end{equation}
\end{widetext}
The similarity between the r.h.s of \eqref{eq:gaussianop} and \eqref{eq:master} suggests to identify $\mathds{E}_\xi[\ket{\psi_\xi}\bra{\psi_\xi}]$ with the l.h.s of \eqref{eq:gaussianop} with $\hat{a}^k_s = \hat{A}_L^k(s)$, $\hat{b}^k_s=\hat{A}^k_R(s)$ and $C_{ij}(\tau,s) = D_{ij}(\tau,s)$. Actually, this does not fully do the trick, a counter term remains, and one sees following \cite{diosi2014} that the correct prescription is:
\begin{align}\label{eq:propagator}
    \ket{\psi_\xi(t)} =& \hat{G}_\xi(t) \ket{\psi_0} \nonumber\\
    :=&\mathcal{T} \exp\bigg\{-i\int_0^t \!\! \upd s\hat{A}^k(s) \xi_k(s) \\
    &\!\!\!\!-\!\! \int_0^t\!\!\int_0^t \!\! \upd \tau\, \upd s\,  \theta_{\tau s} \,  [D-S]_{ij} (\tau,s) \hat{A}^i(\tau)\hat{A}^j(s)\bigg\}  \ket{\psi_0},\nonumber
\end{align}
which fulfills the unraveling condition \eqref{eq:unraveling}.
Equation \eqref{eq:propagator} can subsequently be written in differential form to yield a non-Markovian stochastic Schr\"odinger equation:
\begin{equation}\label{eq:differential}
\begin{split}
\frac{\upd}{\upd t}\ket{\psi_\xi(t)}=&-i\hat{A}^i(t)\bigg[ \xi_i(t) \\
&+ \int_0^t \!\! \upd s \, [D-S]_{ij}(t,s) \frac{\delta}{\delta \xi_j(s)}\bigg] \ket{\psi_{\xi}(t)}.
\end{split}
\end{equation}
To our knowledge, this equation is the most general \emph{linear} non-Markovian stochastic Schr\"odinger equation ever proposed \cite{diosi2014}. Unfortunately, equation \eqref{eq:differential} is quite formal unless it is possible to write the functional derivative\footnote{The reader puzzled by the rigorous definition of the functional derivative can as well use \eqref{eq:propagator} in all situations.} as a simple local operator acting on $\ket{\psi_\xi(t)}$. 

\section{Main result} 

The core objective of this article is to show how the solutions of \eqref{eq:differential} can nonetheless be computed in the general case by reusing a stochastic unraveling trick, \ie by writing this time:
\begin{equation}\label{eq:pureunraveling}
\ket{\psi_\xi(t)}= \mathds{E}_\eta \big[\, \ket{\psi_{\xi,\eta}(t)}\,\big]
\end{equation}
where $\eta$ is an auxiliary set of classical Gaussian fields. More precisely, we can try to write the evolution operator of \eqref{eq:propagator} in the following way:
\begin{equation}\label{eq:localpropagator}
\hat{G}_\xi(t) = \mathds{E}_{\eta}\left[\mathcal{T} \exp \left\{-i \! \int_0^t\!\! \upd s \hat{A}^k(s) [\xi_k(s) + \eta_k(s)]\right\}\right].
\end{equation}
Using again the Gaussian integration formula \eqref{eq:gaussianop}, one sees that the new unraveling condition \eqref{eq:pureunraveling} is satisfied provided $\eta$ is a set of Gaussian fields with zero mean and two-point correlation functions given by:
\begin{equation}\label{eq:correleta}
\begin{split}
\mathds{E}_\eta[\eta_i(\tau)\, \eta_j(s)]&=K_{ij}(\tau,s)\\
\mathds{E}_\eta[\eta_i(\tau)\, \eta^*_j(s)]&=J_{ij}(\tau,s)
\end{split}
\end{equation}
with:
\begin{equation}\label{eq:k}
K_{ij}(\tau,s) = \theta_{\tau s} \,  [D-S]_{ij} (\tau,s) +\theta_{s \tau} \,  [D-S]_{ji} (s,\tau)
\end{equation}
and where the other correlation function $J$ can be chosen freely provided the total kernel $\Gamma$:
\begin{equation}
\Gamma=\left(\begin{array}{cc}
J& K\\
K^*& J^*
\end{array}\right)
\end{equation}
is positive semi-definite\footnote{For a kernel $K$ given by equation \eqref{eq:k}, there always exists a kernel $J$ such that $\Gamma$ is positive semi-definite. The existence of a random field $\eta$ verifying \eqref{eq:correleta} is then a consequence of the Bochner-Minlos theorem.}.

Equation \eqref{eq:localpropagator} can be written in an explicit linear differential form free of functional derivatives:
\begin{equation}\label{eq:diffeta}
\frac{\upd}{\upd t}\ket{\psi_{\xi,\eta}(t)} = - i\hat{A}^k(t) [\xi_k(t) + \eta_k(t)] \, \ket{\psi_{\xi,\eta}(t)}.
\end{equation}
This latter equation, together with its unraveling interpretation \eqref{eq:pureunraveling} is the main result of this article. The evolution given by equation \eqref{eq:diffeta} is \emph{time-local} in the sense that once the random fields $\xi$ and $\eta$ are fixed, the state can be evolved without reference to the past. Although our interest was primarily in the stochastic pure state $\ket{\psi_\xi}$ itself, we can write the open-system density matrix as a double average over our successive unravelings \eqref{eq:unraveling} and \eqref{eq:pureunraveling}:
\begin{equation}
\rho(t) = \mathds{E}_\xi \Big[\, \mathds{E}_\eta \big[\, \ket{\psi_{\xi,\eta}(t)}\,\big]\; \, \mathds{E}_\eta\big[\,\bra{\psi_{\xi,\eta}(t)}\,\big]\,\Big].
\end{equation}
This can be written equivalently:
\begin{equation}
\rho(t) = \mathds{E}_{\xi,\eta^{(1)},\eta^{(2)}} \big[\, \ket{\psi_{\xi,\eta^{(1)}}(t)} \bra{\psi_{\xi,\eta^{(2)}}(t)}\, \big],
\end{equation}
where $\eta^{(1)}$ and $\eta^{(2)}$ are independent. This is the two state unraveling proposal of Stockburger and Grabert \cite{stockburger2002,stockburger2004}. We have thus found a connection between the standard stochastic pure state representations and the more exotic two state vector method used in numerical approaches to open-systems. This allows us to identify at least part of the freedom in the noise present in this latter method as coming from the kernel $S$ which encodes different stochastic Schr\"odinger evolutions corresponding to the same open-system evolution. We can further risk the following heuristic interpretation of the two noises. The noise $\xi$ is classical in the sense that it is averaged over at the density matrix level. The noise $\eta$, on the other hand, is quantum or coherent as all the possible contributions are summed over coherently at the pure state level.

\section{Norm preserving unravelings}

\subsection{Context and objectives}
The linear stochastic differential equation \eqref{eq:differential} does not preserve the norm of the state vector $\ket{\psi_\xi}$. Normalized pure states are usually preferred in foundations but also for numerics: the norm of $\ket{\psi_\xi}$ typically diffuses and for large times, most trajectories give a vanishing contribution to the average \eqref{eq:unraveling} while nearly all the weight is provided by rare events. 

To obtain a norm preserving evolution from the linear one, one needs to define a normalized state $\ket{\widetilde{\psi}_\xi}$:
\begin{equation}
 \ket{\widetilde{\psi}_\xi(t)}=\frac{ \ket{\psi_\xi(t)}}{\sqrt{\langle\psi_\xi(t)|\psi_\xi(t) \rangle}}
\end{equation}
and also to introduce a new ``cooked'' probability measure:
\begin{equation}
 \upd \mu_t(\xi) = \langle\psi_\xi(t)|\psi_\xi(t) \rangle \,\upd \mu_\circ(\xi)
\end{equation}
which insures that the unraveling condition $\rho(t)= \mathds{E}^{t}_\xi\left[\, \ket{\widetilde{\psi}_\xi(t)} \bra{\widetilde{\psi}_\xi(t)}\right]$ --where $\mathds{E}^t_\xi$ is the expectation value taken with the new measure $\mu_t$-- is still valid. It is important to note at that stage that the state $\ket{\widetilde{\psi}_{\xi}(t)}$ with the probability measure $\upd \mu_t(\xi)$ unravels the open-system evolution only for a single instant $t$. If one wants a trajectory that unravels the open-system evolution at all times, it is necessary to dynamically redefine the field variable $\xi$. It can be done by introducing $\xi^{[t]}$, a \emph{complete} random field trajectory depending on $\xi$ such that for any functional $f$:
\begin{equation}\label{eq:variablechange}
\mathds{E}^t_\xi[f(\xi)] = \mathds{E}_\xi \Big[f\big(\xi^{[t]}(\xi)\big)\Big].
\end{equation}
This way, starting from a field realization $\xi$ drawn from the Gaussian measure $\upd \mu_\circ$, one can compute a trajectory $t\mapsto \ket{\widetilde{\psi}_{\xi^{[t]}}(t)}$ that unravels the open evolution for \emph{all} times. By ``computing the norm preserving unraveling'' one may thus want to sample two different things:
\begin{enumerate}
\item the state $\ket{\widetilde{\psi}_{\xi^{[t]}}(t)}$ for a single time,
\item the full trajectory $t \mapsto \ket{\widetilde{\psi}_{\xi^{[t]}}(t)}$. 
\end{enumerate}
While knowing the state at a single time may be enough in most cases, it is not sufficient to compute correlation functions of the stochastic states at different times (something one might want to do for collapse models in foundations). In any case, the ideal solution would be to find a way to write an explicit stochastic differential equation sampling the states of interest directly. This is unfortunately not a trivial endeavour and the method we shall propose is arguably less appealing than in the linear case. 

\subsection{Computation of the full trajectory for \texorpdfstring{S=0}{}}\label{sec:nonlinear}
When $S=0$, there exists an efficient way to construct the full norm preserving trajectory $t \mapsto \ket{\widetilde{\psi}_{\xi^{[t]}}(t)}$, exploiting the fact that $\xi$ and $\xi^{[t]}$, defined implicitly in \eqref{eq:variablechange}, can be connected explicitly in that case. Indeed, after relatively painful computations (see \cite{gisin1992,wiseman1993,tilloy2017}) one can show that for $S=0$:
\begin{equation}\label{eq:transform}
\xi^{[t]}_k(u) = \xi_k(u) + i\int_0^t \upd s\,D_{k\ell}(u,s) \langle A^\ell(s)\rangle_s,
\end{equation}
with $\langle A_\ell(s)\rangle_s=\bra{\widetilde{\psi}_{\xi^{[s]}}(s)}A^\ell(s)\ket{\widetilde{\psi}_{\xi^{[s]}}(s)}$. Using this result, it is possible to compute $\ket{\widetilde{\psi}_{\xi^{[t]}}(t)}$ inductively. 

\begin{figure}
\centering
\includegraphics[width=0.99\columnwidth]{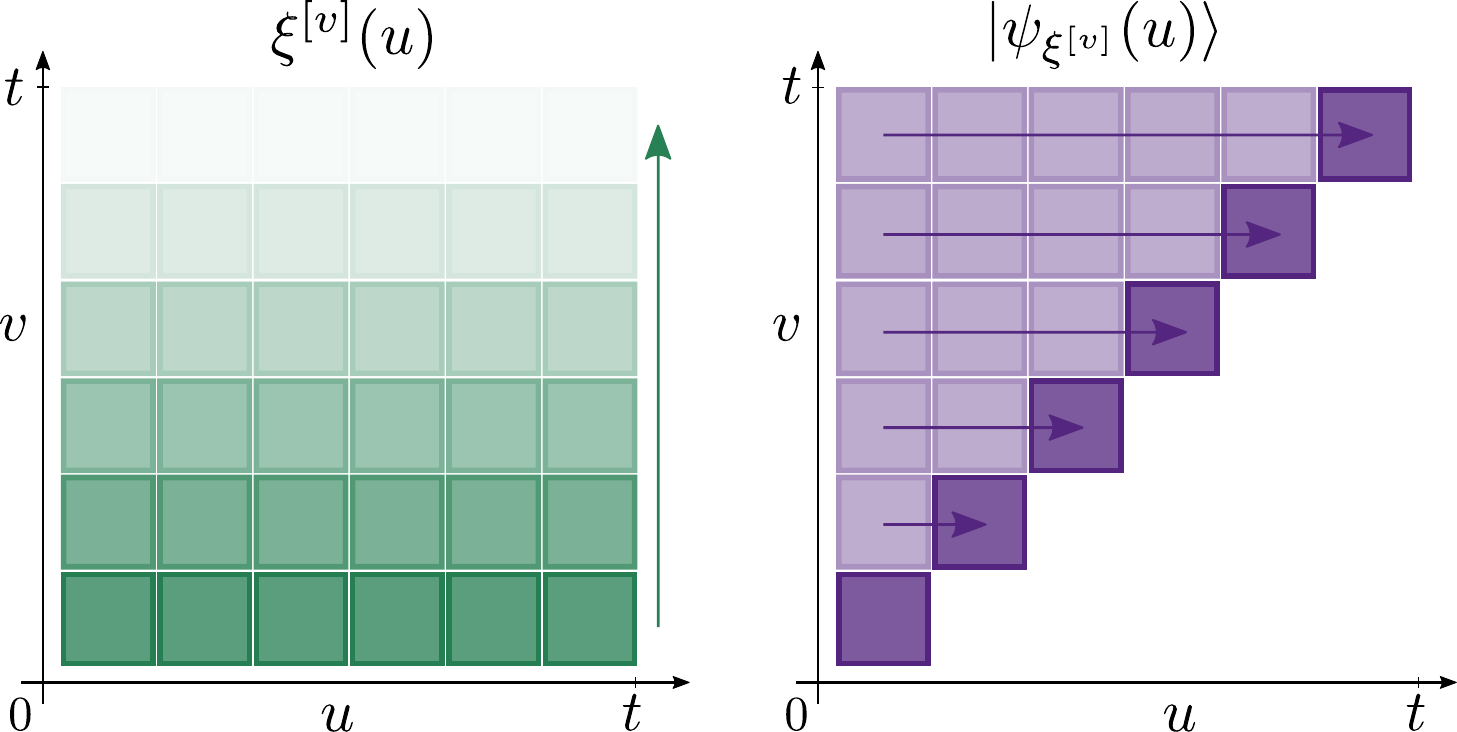}
\caption{Schematic representation of the procedure described in \ref{sec:nonlinear}. The states and fields are computed from bottom to top for increasing values of $v$. The final product, the norm-preserving trajectory, is obtained for $u=v$.}
\label{fig:explanation}
\end{figure}

We start from $\xi^{[0]}=\xi$ sampled from the Gaussian measure $\upd \mu_\circ$. At a time $v\leq t$ we assume that:
\begin{itemize}
    \item[--]$\forall u\leq v$, $\ket{\widetilde{\psi}_{\xi^{[u]}}(u)}$ is known,
    \item[--]$\forall u \leq t$, $\xi^{[v]}_u$ is known.
\end{itemize}
We first compute $\xi^{[v+\upd v]}$ for $u\leq t$ using \eqref{eq:transform}:
\begin{equation}\label{eq:discrete_update}
\xi^{[v+\upd v]}_k(u) = \xi^{[v]}_k(u) + iD_{k,\ell}(u,v) \langle A^\ell(v)\rangle_v \upd v + O(\upd v^2),
\end{equation}
We then compute $\ket{\widetilde{\psi}_{\xi^{[v+\upd v]}}(v+\upd v)}$ solving the linear equation \eqref{eq:differential} up to time $v+\upd v$ with the new complete noise trajectory $\xi^{[v+\upd v]}$ and normalizing the result. This is done using the unraveling \eqref{eq:diffeta}. By induction, we thus obtain $\ket{\widetilde{\psi}_{\xi^{[t]}}(t)}$ for all $t$ starting from a single realization of the noise $\xi$ (see figure \ref{fig:explanation}). For that matter, it is necessary to solve the linear stochastic Schr\"odinger equation from the beginning with a new field at every time step. Hence this adds a linear a overhead $\propto t/\upd t$ to the time needed to compute linear trajectories. As the errors coming from the discretization are typically much smaller than the statistical errors coming from the averaging of \eqref{eq:diffeta}, there is however no loss in precision.

\begin{figure*}
\centering
\includegraphics[width=0.49\textwidth]{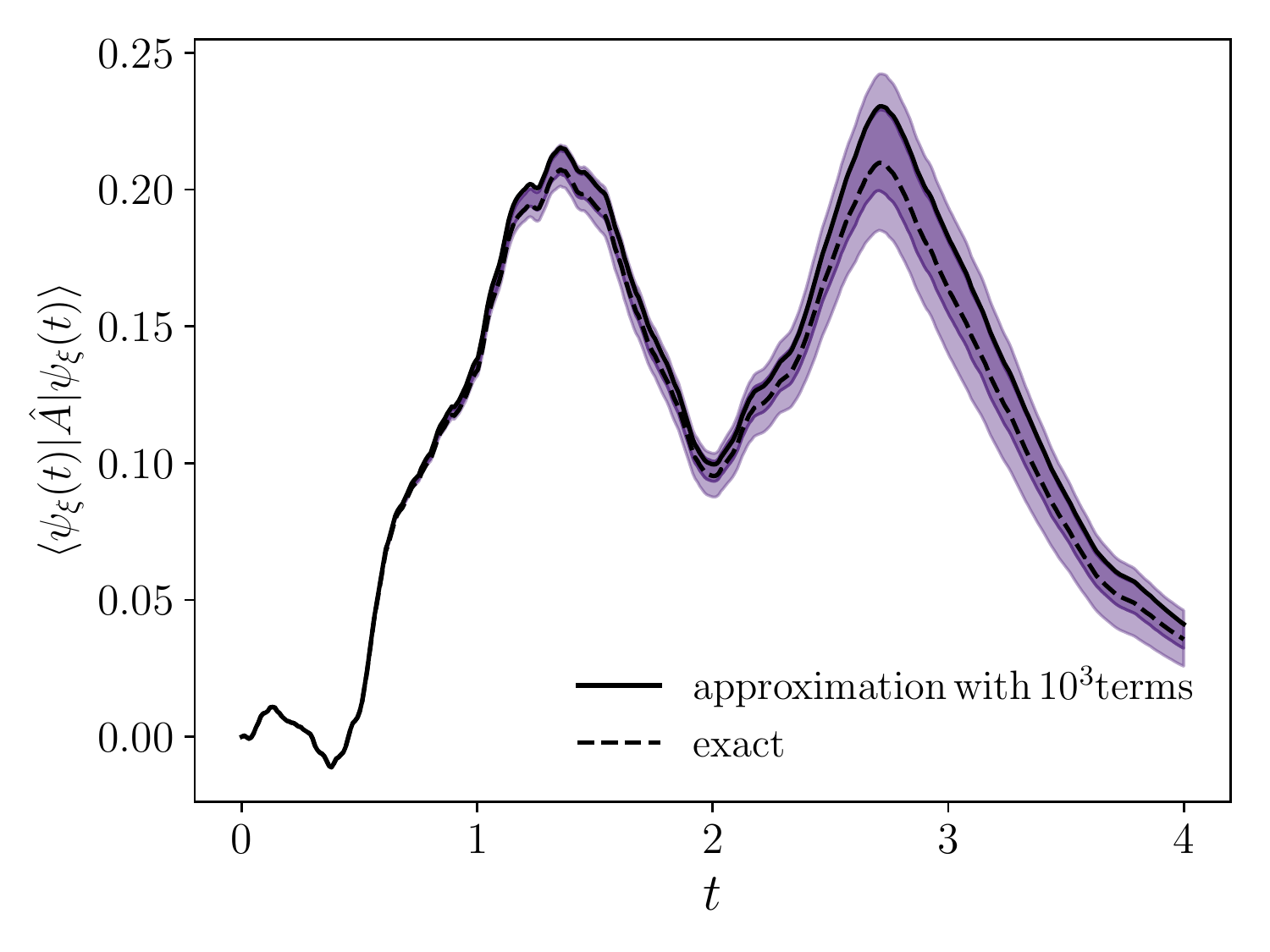}
\includegraphics[width=0.49\textwidth]{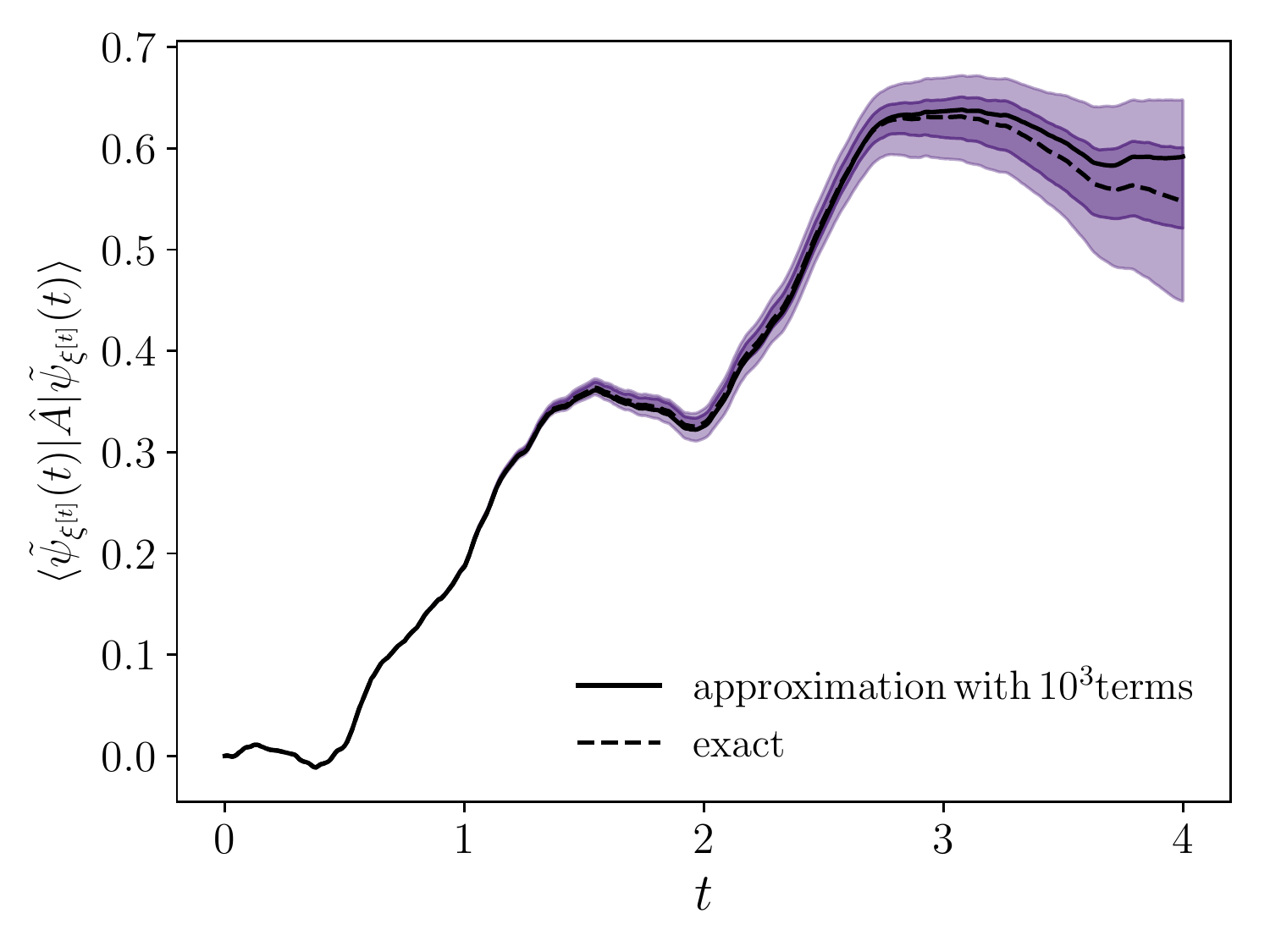}
\caption{\textbf{Left:} Snapshot of a linear trajectory $\ket{\psi_\xi}$ for a random realization of $\xi$.  The exact trajectory is obtained from equation \eqref{eq:exact} while the approximate trajectory is obtained from \eqref{eq:diffeta} --integrated with $\upd t=10^{-2}$--, and the estimation $\ket{\psi_\xi}\simeq\frac{1}{N}\sum_{n=1}^{N}\ket{\psi_{\xi,\eta}}$ with $N=10^3$ independent realizations of $\eta$ and the same fixed $\xi$. 
\textbf{Right:} Snapshot of a non-linear trajectory $\ket{\widetilde{\psi}_{\xi^{[t]}}(t)}$ for the same realization $\xi$, and computed with the method of \ref{sec:nonlinear}. In the algorithm, the linear NMSSE is solved by integrating \eqref{eq:diffeta} with  $\upd t=10^{-2}$ and averaging over $N=10^3$ independent realizations of $\eta$.\\
The shaded areas indicate the zones where $50\%$ ($\simeq 0.5\sigma$) and $85\%$ ($\simeq 1 \sigma$) of the approximate trajectories would lie (obtained by computing $100$ samples for the same $\xi$). In both cases, number of realizations is voluntarily left small to make the estimate of the error visible as the latter decreases $\propto N^{-1/2}$.}
\label{fig:snapshots}
\end{figure*}
\subsection{General case, single time}\label{sec:single_time}

In the general case, the previous method cannot be used as we do not know of a generalization of formula \eqref{eq:variablechange} for $S\neq0$. Finding a way to sample from norm-preserving trajectories has eluded us. In most cases however, one is only interested in the properties of the stochastic state at a single time. In such a situation, the linear unraveling is sufficient to compute everything. Indeed, let us consider an arbitrary function $\varphi$ of the norm-preserving trajectory $\ket{\widetilde{\psi}_{\xi^{[t]}}(t)}$ at a \emph{fixed} time $t$. We have:
\begin{align}
\mathds{E}&\big[\varphi(\ket{\widetilde{\psi}_{\xi^{[t]}}(t)})\big]=\mathds{E}^t_\xi\big[\varphi(\ket{\widetilde{\psi}_{\xi}(t)})\big]\\
&=\mathds{E}^t_\xi\left[\varphi\left(\frac{\ket{\psi_{\xi}(t)}}{\sqrt{\langle\psi_{\xi}(t) | \psi_{\xi}(t)\rangle}}\right)  \right]\\
&=\mathds{E}_\xi\left[\varphi\left(\frac{\ket{\psi_{\xi}(t)}}{\sqrt{\langle\psi_{\xi}(t) | \psi_{\xi}(t)\rangle}}\right)  \langle\psi_{\xi}(t) | \psi_{\xi}(t)\rangle\right].
\end{align}
That is, expectation values of all time-local functionals of the norm preserving trajectory can be computed by averaging over the solutions of the linear NMSSE.

\section{Example}

\subsection{Linear evolution}

We now illustrate our method on an example where the stochastic Schr\"odinger equation is fully explicit.
For that matter, we consider a single Hermitian operator $\hat{A}$ commuting with $H_0$ (s.t. $\hat{A}(t)=\hat{A}$) and purely isotropic noise, \ie $S=0$, which gives the following stochastic Schr\"odinger equation:
\begin{equation}\label{eq:exact}
 \frac{\upd}{\upd t} \ket{\psi_\xi(t)}=-i\hat{A} \left[\xi(t) + \int_0^t \!\! \upd s \, D(t,s)\frac{\delta}{\delta \xi(s)}\right]\! \ket{\psi_\xi(t)}.
\end{equation}
For a given realization of $\xi$, $\ket{\psi_\xi}$ obeys an explicit differential equation. Indeed, using the integral form \eqref{eq:propagator}, we can replace the functional derivative:
\begin{equation}
\frac{\delta}{\delta \xi(s)} \ket{\psi_\xi(t)}= -i\hat{A}\ket{\psi_\xi(t)}.
\end{equation}
This example thus offers a nice test bed for our stochastic approach. In this example, the relaxation timescale responsible for the non-Markovianity and the timescale of the system-bath coupling are both of order $1$ and we would thus be deep in the non-Markovian regime had the system Hamiltonian not been trivial. To compute the linear trajectories, we average over the solutions of \eqref{eq:diffeta} with $\eta$ a real noise such that $\mathds{E}[\eta_t \eta_s]=D(t,s)$.  Numerical results for $\hat{A}=\mathrm{diag} (1,0,-1)$, $D(t,0)=e^{- t}$, and $\ket{\psi(0)}=(1,1,1)/\sqrt{3}$ are shown in figure \ref{fig:snapshots}.

Our method is efficient for evolution times of the order of the the system-bath coupling timescale but becomes expensive for larger times. Indeed, we are summing states $\ket{\psi_{\xi,\eta}(t)}$ with a phase with respect to $\ket{\psi_\xi(t)}$ that is more and more random as time grows. This is an important limit of this scheme for large time.

\subsection{Norm preserving evolution}
We also sample the norm-preserving non-linear trajectories using the method of section \ref{sec:nonlinear}. The precision for the sampling of a single trajectory $\ket{\psi_{\xi^{[t]}}(t)}$ is very similar to that of the linear case (see figure \ref{fig:snapshots}).

\begin{figure}
\centering
\includegraphics[width=0.99\columnwidth]{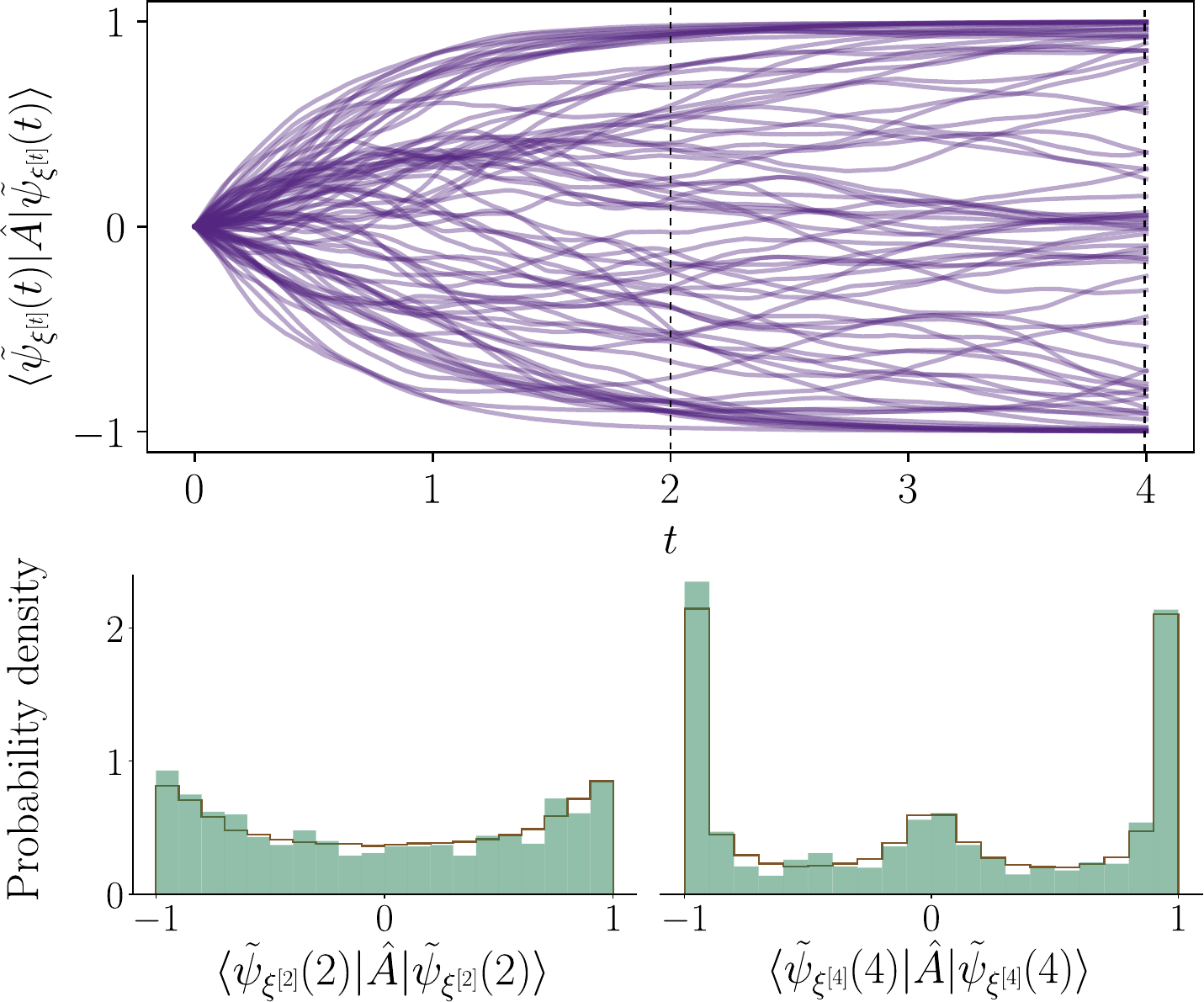}
\caption{\textbf{Top:} Snapshot of $80$ normalized trajectories computed with the method of \ref{sec:nonlinear}. In the algorithm, the linear equation is solved by integrating \eqref{eq:diffeta} with  $\upd t=10^{-2}$ and averaging over $N=10^3$ independent realizations of $\eta$.\\
\textbf{Bottom:} Empirical probability densities; green histogram computed by sampling $10^3$ non-linear trajectories (each computed with the same method as in the snapshot), brown line computed by sampling $10^5$ \emph{exact} linear trajectories and re-weighting a posteriori as in \ref{sec:single_time}}
\label{fig:histo}
\end{figure}

To test the robustness of the method, we also compute the probability distribution of $\bra{\psi_{\xi^{[t]}}(t)} \hat{A}\ket{\psi_{\xi^{[t]}}(t)}$ for two different times using norm-preserving trajectories computed with the numerical method of \ref{sec:nonlinear} and using exact linear trajectories and the re-weighting of \ref{sec:single_time}. Up to statistical and discretization errors, the second method is exact. The results are displayed in figure \ref{fig:histo} and show an excellent agreement between the two methods.

\section{Conclusion}
In this article, we have introduced a rewriting (or ``unraveling'') of the solutions of formal NMSSE as averages over the solutions of explicit time-local stochastic differential equations. This new formulation may provide a basis for further analytical studies, notably as it gives an extremely simple new way to define NMSSE. More importantly, it immediately offers a method to compute numerically their solutions. 
Although we have proposed a heuristic interpretation of the quantum and classical character of the two noises involved, the auxiliary state $\ket{\psi_{\xi,\eta}}$ does not yet have a clear operational characterization. The latter would certainly help understand the freedom in the noise kernel $J$.
We have also extended our method to non-linear norm-preserving NMSSE in the case where $S=0$, \ie where the complex noise $\xi$ is isotropic. Computing expectation values of time non-local functionals of norm-preserving trajectories when $S\neq 0$ is still an open problem.
Our work has been focused on the NMSSE as an end in themselves rather than on the open-system evolution they unravel. It is however possible that our methods, especially the one to sample non-linear NMSSE directly, might ultimately help improve our ability to deal with non-Markovian open-systems numerically.

\begin{acknowledgments}
I thank Lajos Di\'osi and Luca Ferialdi for helpful discussions. Part of this work was supported by the Alexander von Humboldt foundation and the Agence
Nationale de la Recherche (ANR) contract ANR-14-CE25-0003-01.
\end{acknowledgments}

\bibliographystyle{plainnat_trimmed}
\bibliography{main}

\begin{thebibliography}{37}
\providecommand{\natexlab}[1]{#1}
\providecommand{\url}[1]{\texttt{#1}}
\expandafter\ifx\csname urlstyle\endcsname\relax
  \providecommand{\doi}[1]{doi: #1}\else
  \providecommand{\doi}{doi: \begingroup \urlstyle{rm}\Url}\fi

\bibitem[Adler and Bassi(2007)]{adler2007}
S.~L. Adler and A. Bassi.
\newblock Collapse models with non-white noises.
\newblock \emph{Journal of Physics A: Mathematical and Theoretical},
  40\penalty0 (50):\penalty0 15083, 2007.
\newblock \doi{10.1088/1751-8113/40/50/012}.

\bibitem[Bassi and Ghirardi(2003)]{bassi2003}
A. Bassi and G. Ghirardi.
\newblock Dynamical reduction models.
\newblock \emph{Physics Reports}, 379\penalty0 (5–6):\penalty0 257 -- 426,
  2003.
\newblock \doi{10.1016/S0370-1573(03)00103-0}.

\bibitem[Bassi et~al.(2013)Bassi, Lochan, Satin, Singh, and
  Ulbricht]{bassi2013}
A. Bassi, K. Lochan, S. Satin, T.~P. Singh, and H. Ulbricht.
\newblock Models of wave-function collapse, underlying theories, and
  experimental tests.
\newblock \emph{Rev. Mod. Phys.}, 85:\penalty0 471--527, Apr 2013.
\newblock \doi{10.1103/RevModPhys.85.471}.

\bibitem[Dalibard et~al.(1992)Dalibard, Castin, and M\o{}lmer]{dalibard1992}
J. Dalibard, Y. Castin, and K. M\o{}lmer.
\newblock Wave-function approach to dissipative processes in quantum optics.
\newblock \emph{Phys. Rev. Lett.}, 68:\penalty0 580--583, Feb 1992.
\newblock \doi{10.1103/PhysRevLett.68.580}.

\bibitem[de~Vega et~al.(2005)de~Vega, Alonso, and Gaspard]{devega2005}
I. de~Vega, D. Alonso, and P. Gaspard.
\newblock Two-level system immersed in a photonic band-gap material: A
  non-markovian stochastic schr\"odinger-equation approach.
\newblock \emph{Phys. Rev. A}, 71:\penalty0 023812, Feb 2005.
\newblock \doi{10.1103/PhysRevA.71.023812}.

\bibitem[Di\'osi(1989)]{diosi1989}
L. Di\'osi.
\newblock Models for universal reduction of macroscopic quantum fluctuations.
\newblock \emph{Phys. Rev. A}, 40:\penalty0 1165--1174, Aug 1989.
\newblock \doi{10.1103/PhysRevA.40.1165}.

\bibitem[Di\'osi and Ferialdi(2014)]{diosi2014}
L. Di\'osi and L. Ferialdi.
\newblock General non-markovian structure of gaussian master and stochastic
  schr\"odinger equations.
\newblock \emph{Phys. Rev. Lett.}, 113:\penalty0 200403, Nov 2014.
\newblock \doi{10.1103/PhysRevLett.113.200403}.

\bibitem[Di\'osi(1986)]{diosi1986}
L. Di\'osi.
\newblock Stochastic pure state representation for open quantum systems.
\newblock \emph{Phys. Lett. A}, 114\penalty0 (8):\penalty0 451 -- 454, 1986.
\newblock \doi{10.1016/0375-9601(86)90692-4}.

\bibitem[Di\'osi(2008{\natexlab{a}})]{diosi2008}
L. Di\'osi.
\newblock Non-markovian continuous quantum measurement of retarded observables.
\newblock \emph{Phys. Rev. Lett.}, 100:\penalty0 080401, Feb
  2008{\natexlab{a}}.
\newblock \doi{10.1103/PhysRevLett.100.080401}.

\bibitem[Di\'osi(2008{\natexlab{b}})]{diosi2008erratum}
L. Di\'osi.
\newblock Erratum: Non-markovian continuous quantum measurement of retarded
  observables [phys. rev. lett. \textbf{100} , 080401 (2008)].
\newblock \emph{Phys. Rev. Lett.}, 101:\penalty0 149902, Oct
  2008{\natexlab{b}}.
\newblock \doi{10.1103/PhysRevLett.101.149902}.

\bibitem[Di{\'o}si and Strunz(1997)]{diosi1997}
L. Di{\'o}si and W.~T. Strunz.
\newblock The non-markovian stochastic schr{\"o}dinger equation for open
  systems.
\newblock \emph{Phys. Lett. A}, 235\penalty0 (6):\penalty0 569--573, 1997.
\newblock \doi{10.1016/S0375-9601(97)00717-2}.

\bibitem[Dum et~al.(1992)Dum, Zoller, and Ritsch]{dum1992}
R. Dum, P. Zoller, and H. Ritsch.
\newblock Monte carlo simulation of the atomic master equation for spontaneous
  emission.
\newblock \emph{Phys. Rev. A}, 45:\penalty0 4879--4887, Apr 1992.
\newblock \doi{10.1103/PhysRevA.45.4879}.

\bibitem[Ferialdi and Bassi(2012)]{ferialdi2012}
L. Ferialdi and A. Bassi.
\newblock Dissipative collapse models with nonwhite noises.
\newblock \emph{Phys. Rev. A}, 86:\penalty0 022108, Aug 2012.
\newblock \doi{10.1103/PhysRevA.86.022108}.

\bibitem[Feynman and Vernon(1963)]{feynman1963}
R.~P. Feynman and F.~L. Vernon.
\newblock The theory of a general quantum system interacting with a linear
  dissipative system.
\newblock \emph{Annals of physics}, 24:\penalty0 118--173, 1963.
\newblock \doi{10.1016/0003-4916(63)90068-X}.

\bibitem[Gambetta and Wiseman(2002)]{gambetta2002}
J. Gambetta and H.~M. Wiseman.
\newblock Non-markovian stochastic schr\"odinger equations: Generalization to
  real-valued noise using quantum-measurement theory.
\newblock \emph{Phys. Rev. A}, 66:\penalty0 012108, Jul 2002.
\newblock \doi{10.1103/PhysRevA.66.012108}.

\bibitem[Gambetta and Wiseman(2003)]{gambetta2003}
J. Gambetta and H.~M. Wiseman.
\newblock Interpretation of non-markovian stochastic schr\"odinger equations as
  a hidden-variable theory.
\newblock \emph{Phys. Rev. A}, 68:\penalty0 062104, Dec 2003.
\newblock \doi{10.1103/PhysRevA.68.062104}.

\bibitem[Ghirardi et~al.(1990)Ghirardi, Pearle, and Rimini]{ghirardi1990}
G.~C. Ghirardi, P. Pearle, and A. Rimini.
\newblock Markov processes in hilbert space and continuous spontaneous
  localization of systems of identical particles.
\newblock \emph{Phys. Rev. A}, 42:\penalty0 78--89, Jul 1990.
\newblock \doi{10.1103/PhysRevA.42.78}.

\bibitem[Gisin(1984)]{gisin1984}
N. Gisin.
\newblock Quantum measurements and stochastic processes.
\newblock \emph{Phys. Rev. Lett.}, 52:\penalty0 1657--1660, May 1984.
\newblock \doi{10.1103/PhysRevLett.52.1657}.

\bibitem[Gisin and Percival(1992)]{gisin1992}
N. Gisin and I.~C. Percival.
\newblock The quantum-state diffusion model applied to open systems.
\newblock \emph{J. Phys. A: Math. Gen.}, 25\penalty0 (21):\penalty0 5677, 1992.
\newblock \doi{10.1088/0305-4470/25/21/023}.

\bibitem[Hu et~al.(1992)Hu, Paz, and Zhang]{hu1992}
B.~L. Hu, J.~P. Paz, and Y. Zhang.
\newblock Quantum brownian motion in a general environment: Exact master
  equation with nonlocal dissipation and colored noise.
\newblock \emph{Phys. Rev. D}, 45:\penalty0 2843--2861, Apr 1992.
\newblock \doi{10.1103/PhysRevD.45.2843}.

\bibitem[Hu et~al.(1993)Hu, Paz, and Zhang]{hu1993}
B.~L. Hu, J.~P. Paz, and Y. Zhang.
\newblock Quantum brownian motion in a general environment. ii. nonlinear
  coupling and perturbative approach.
\newblock \emph{Phys. Rev. D}, 47:\penalty0 1576--1594, Feb 1993.
\newblock \doi{10.1103/PhysRevD.47.1576}.

\bibitem[Jacobs and Steck(2006)]{jacobs2006}
K. Jacobs and D.~A. Steck.
\newblock A straightforward introduction to continuous quantum measurement.
\newblock \emph{Contemporary Physics}, 47\penalty0 (5):\penalty0 279--303,
  2006.
\newblock \doi{10.1080/00107510601101934}.

\bibitem[Jing and Yu(2010)]{jing2010}
J. Jing and T. Yu.
\newblock Non-markovian relaxation of a three-level system: Quantum trajectory
  approach.
\newblock \emph{Phys. Rev. Lett.}, 105:\penalty0 240403, Dec 2010.
\newblock \doi{10.1103/PhysRevLett.105.240403}.

\bibitem[Jing et~al.(2012)Jing, Zhao, You, and Yu]{jing2012}
J. Jing, X. Zhao, J.~Q. You, and T. Yu.
\newblock Time-local quantum-state-diffusion equation for multilevel quantum
  systems.
\newblock \emph{Phys. Rev. A}, 85:\penalty0 042106, Apr 2012.
\newblock \doi{10.1103/PhysRevA.85.042106}.

\bibitem[Jing et~al.(2013)Jing, Zhao, You, Strunz, and Yu]{jing2013}
J. Jing, X. Zhao, J.~Q. You, W.~T. Strunz, and T. Yu.
\newblock Many-body quantum trajectories of non-markovian open systems.
\newblock \emph{Phys. Rev. A}, 88:\penalty0 052122, Nov 2013.
\newblock \doi{10.1103/PhysRevA.88.052122}.

\bibitem[M{\o}lmer et~al.(1993)M{\o}lmer, Castin, and Dalibard]{molmer1993}
K. M{\o}lmer, Y. Castin, and J. Dalibard.
\newblock Monte carlo wave-function method in quantum optics.
\newblock \emph{J. Opt. Soc. Am. B}, 10\penalty0 (3):\penalty0 524--538, Mar
  1993.
\newblock \doi{10.1364/JOSAB.10.000524}.

\bibitem[Pearle(1989)]{pearle1989}
P. Pearle.
\newblock Combining stochastic dynamical state-vector reduction with
  spontaneous localization.
\newblock \emph{Phys. Rev. A}, 39:\penalty0 2277--2289, Mar 1989.
\newblock \doi{10.1103/PhysRevA.39.2277}.

\bibitem[Stockburger(2004)]{stockburger2004}
J.~T. Stockburger.
\newblock Simulating spin-boson dynamics with stochastic liouville--von neumann
  equations.
\newblock \emph{Chemical physics}, 296\penalty0 (2):\penalty0 159--169, 2004.
\newblock \doi{10.1016/j.chemphys.2003.09.014}.

\bibitem[Stockburger and Grabert(2002)]{stockburger2002}
J.~T. Stockburger and H. Grabert.
\newblock Exact $\mathit{c}$-number representation of non-markovian quantum
  dissipation.
\newblock \emph{Phys. Rev. Lett.}, 88:\penalty0 170407, Apr 2002.
\newblock \doi{10.1103/PhysRevLett.88.170407}.

\bibitem[Strunz(1996)]{strunz1996}
W.~T. Strunz.
\newblock Linear quantum state diffusion for non-markovian open quantum
  systems.
\newblock \emph{Phys. Lett. A}, 224\penalty0 (1):\penalty0 25 -- 30, 1996.
\newblock \doi{10.1016/S0375-9601(96)00805-5}.

\bibitem[Strunz and Yu(2004)]{strunz2004}
W.~T. Strunz and T. Yu.
\newblock Convolutionless non-markovian master equations and quantum
  trajectories: Brownian motion.
\newblock \emph{Phys. Rev. A}, 69:\penalty0 052115, May 2004.
\newblock \doi{10.1103/PhysRevA.69.052115}.

\bibitem[Strunz et~al.(1999)Strunz, Di\'osi, Gisin, and Yu]{strunz1999}
W.~T. Strunz, L. Di\'osi, N. Gisin, and T. Yu.
\newblock Quantum trajectories for brownian motion.
\newblock \emph{Phys. Rev. Lett.}, 83:\penalty0 4909--4913, Dec 1999.
\newblock \doi{10.1103/PhysRevLett.83.4909}.

\bibitem[Tilloy(2017)]{tilloy2017}
A. Tilloy.
\newblock Interacting quantum field theories as relativistic statistical field
  theories of local beables.
\newblock \emph{arXiv:1702.06325}, 2017.

\bibitem[Weiss(1999)]{weiss1999}
U. Weiss.
\newblock \emph{Quantum dissipative systems}, volume~10.
\newblock World Scientific, 1999.

\bibitem[Wiseman and Milburn(1993)]{wiseman1993}
H.~M. Wiseman and G.~J. Milburn.
\newblock Quantum theory of field-quadrature measurements.
\newblock \emph{Phys. Rev. A}, 47:\penalty0 642--662, Jan 1993.
\newblock \doi{10.1103/PhysRevA.47.642}.

\bibitem[Wiseman and Gambetta(2008)]{wiseman2008}
H.~M. Wiseman and J.~M. Gambetta.
\newblock Pure-state quantum trajectories for general non-markovian systems do
  not exist.
\newblock \emph{Phys. Rev. Lett.}, 101:\penalty0 140401, Sep 2008.
\newblock \doi{10.1103/PhysRevLett.101.140401}.

\bibitem[Zhao et~al.(2011)Zhao, Jing, Corn, and Yu]{zhao2011}
X. Zhao, J. Jing, B. Corn, and T. Yu.
\newblock Dynamics of interacting qubits coupled to a common bath:
  Non-markovian quantum-state-diffusion approach.
\newblock \emph{Phys. Rev. A}, 84:\penalty0 032101, Sep 2011.
\newblock \doi{10.1103/PhysRevA.84.032101}.

\end{thebibliography}

\end{document}